\documentstyle[psfig,macros]{mn}
\begin{document}
%

\title[Statistics of Dark Matter Substructure II]
      {Statistics of Dark Matter Substructure:  
       II. Comparison of Model with Simulation Results}

\author[van den Bosch \& Jiang]
       {Frank C. van den Bosch$^{1}$\thanks{E-mail:frank.vandenbosch@yale.edu},
        Fangzhou Jiang$^{1}$\\
          \vspace*{3pt} \\
        $^1$Department of Astronomy, Yale University, PO Box 208101, 
            New Haven, CT 06520-8101}


\date{}

\pagerange{\pageref{firstpage}--\pageref{lastpage}}
\pubyear{2013}

\maketitle

\label{firstpage}


\begin{abstract}
  We compare subhalo mass and velocity functions obtained from
  different simulations with different subhalo finders among each
  other, and with predictions from the new semi-analytical model of Jiang
  \& van den Bosch (2014). We find that subhalo mass functions (SHMFs)
  obtained using different subhalo finders agree with each other at
  the level of $\sim 20$ percent, but only at the low mass end. At the
  massive end, subhalo finders that identify subhaloes based purely on
  density in configuration space dramatically underpredict the subhalo
  abundances by more than an order of magnitude. These problems are
  much less severe for subhalo velocity functions (SHVFs), indicating
  that they arise from issues related to assigning masses to the
  subhaloes, rather than from detecting them.  Overall the predictions
  from the semi-analytical model are in excellent agreement with simulation
  results obtained using the more advanced subhalo finders that use
  information in six dimensional phase-space. In particular, the model
  accurately reproduces the slope and host-mass-dependent
  normalization of both the subhalo mass and velocity functions. We
  find that the SHMFs and SHVFs have power-law slopes of $0.82$ and
  $2.6$, respectively, significantly shallower than what has been
  claimed in several studies in the literature.
\end{abstract} 




\section{Introduction} 
\label{Sec:Introduction}

Hierarchical structure formation in a $\Lambda$CDM cosmology gives
rise to dark matter haloes with abundant substructure in the form of
self-bound clumps of matter. These subhaloes are the remnants of dark
matter haloes that have been accreted by their host halo over cosmic
time, and that have (thus far) survived tidal destruction.  Dark
matter subhaloes host satellite galaxies, boost the dark matter
annihilation signal, cause tidal heating of stellar streams and disks,
and are responsible for time-delays and flux-ratio anomalies in
gravitational lensing. Hence, characterizing the abundance, spatial
distribution and internal structure of dark matter substructure is
important for a large number of astrophysical applications.

Given the highly non-linear nature of hierarchical structure
formation, the substructure of dark matter haloes is best studied
using high-resolution $N$-body simulations. Ever since the first
numerical simulations reached sufficient resolution to resolve dark
matter subhaloes (Tormen 1997; Ghigna \etal 1998; Tormen, Diaferio \&
Syer 1998; Klypin \etal 1999; Moore \etal 1999) numerous studies have
used $N$-body simulations of ever increasing size and/or numerical
resolution to study their statistics as function of host halo mass,
redshift, cosmology, and other properties of relevance, such as the
formation time of the halo (e.g., Springel \etal 2001, 2008; Stoehr
\etal 2002; Gao \etal 2004; De Lucia \etal 2004; Diemand, Moore \&
Stadel 2004; Gill \etal 2004a,b; Kravtsov \etal 2004; Reed \etal 2005;
Shaw \etal 2007; Giocoli \etal 2008a, 2010; Weinberg \etal 2008;
Angulo \etal 2009; Boylan-Kolchin \etal 2010; Klypin, Trujillo-Gomez
\& Primack 2011; Wu \etal 2013). These studies not only used different
simulation codes, different cosmologies, different numerical
resolutions, and different simulation volumes, but also different
subhalo finders.

To date, more than a dozen different subhalo finders have been used,
all based on some of the following two characteristics of a subhalo:
(i) it is a self-bound, overdense region inside its host halo, and
(ii) it was it's own host halo before it merged into its current host
(see Han \etal 2012). Most halo finders only use the instantaneous
particle positions and velocities to identify subhaloes based on the
first characteristic listed above. Most of these only use the velocity
information to remove unbound particles from a list of candidate
particles identified based on density alone. Examples of such halo
finders are \Subfind (Springel \etal 2001), {\tt SKID} (Stadel 2001),
Bound Density Maximum (\BDM; Klypin \& Holtzman 1997), Amiga Halo
Finder ({\tt AHF}; Knollmann \& Knebe 2009), and Voronoi Bound Zones
({\tt VOBOZ}; Neyrinck, Gnedin \& Hamilton 2005). Others, such as
6-Dimensional Friends-of-Friends ({\tt 6DFOF}; Diemand Kuhlen \& Madau
2006), Hierarchical Structure Finder ({\tt HSF}; Maciejewski \etal
2009) and \Rockstar (Behroozi \etal 2013a), identify (sub)haloes
using the full six-dimensional phase-space information.  Finally,
there are also subhalo finders that make additional use of the second
characteristic listed above by employing the time domain.  Since
subhaloes are remnants of dark matter host haloes, one can identify
the former by tracing the member particles of the latter that remain
part of a self-bound entity. Examples of these are \SURV (Tormen \etal
1998) and the Hierarchical Bound-Tracing algorithm ({\tt HBT}) of Han
\etal (2012). Note that \Rockstar also uses some time-domain
information in its (sub)halo identification, making it the only
subhalo finder that uses information in seven dimensions.

In an era of precision cosmology, in which accurate, percent level
characterization of the abundances of dark matter haloes and subhaloes
is crucial, it is of paramount importance to compare the performance
of all these different subhalo finders, and to quantify their accuracy
and reliability. Unfortunately, and somewhat surprisingly, this has
received relatively little attention.  Muldrew, Pearce \& Power (2011)
compared the performances of \Subfind and {\tt AHF} in recovering mock
subhaloes placed in a mock host halo.  They showed that the mass of
the subhalo recovered by \Subfind has a strong dependence on its
radial position within the host halo, and that neither subhalo finder
can accurately recover the subhalo mass when it is near the center of
the host halo. More quantitatively, \Subfind was only able to recover
50 percent of the subhalo mass when its center is located at half the
virial radius from the center of its host. At $r < r_{\rm vir}/10$,
neither subhalo finder could recover more than 40 percent of the
actual subhalo mass. These problems arise from the subhalo being
defined as a mere overdensity in configuration space. Indeed, using a
similar approach based on mock haloes, Knebe \etal (2011) showed that
this problem can be significantly reduced (but not eliminated) when
using a subhalo finder that operates in 6D phase-space. A potential
problem with these two studies, though, is that they used mock
haloes. As pointed out in Knebe \etal (2011), the discrepancy between
the true and recovered subhalo masses is likely to be overestimated in
this idealized setup. In reality, a subhalo experiences tidal
stripping and truncation when moving towards the center of its host
halo, and the mass discrepancy is likely to be reduced when only
considering the mass within the tidal truncation radius. Following up
on the initial study by Knebe \etal (2011), Onions \etal (2012)
therefore compared the performance of subhalo finders using an
ultra-high resolution simulation of a single Milky-Way sized dark
matter halo from the Aquarius project (Springel \etal 2008).
Comparing the statistics and properties of the dark matter subhaloes
identified within this single host halo with ten different subhalo
finders, and using a common post-processing pipeline to uniformly
analyze the particle lists provided by each finder, they find that the
basic properties (mass and maximum circular velocity) of dark matter
subhaloes can be reliably recovered (at an accuracy better than 20
percent) if the subhaloes contain more than 100 particles.  In a
follow-up study, Knebe \etal (2013) showed that discarding the results
from the two subhalo finders that lack a (reliable) procedure to
remove unbound particles, the scatter among the (eight remaining)
subhalo finders is reduced by a factor two, to $\sim 10\%$.  Finally,
the studies of Onions \etal (2012) and Knebe \etal (2013) show that
configuration finders yield less reliable masses for subhaloes close
to the center of their host than phase-space finders, but that the
differences are significantly smaller than suggested by the tests
based on mock haloes described above.

Unfortunately, since the study by Onions et al. only used a single
dark matter halo, albeit at exquisite numerical resolution, the
comparison is limited to the relatively low mass end of the subhalo
mass function, where the cumulative mass function $N(>m)$, exceeds
unity.  In order to study the massive end of the subhalo mass
function, where $N(>m) < 1$, one needs to average over large numbers
of host haloes. The abundances of these rare but massive subhaloes has
important implications for, among others, the statistics of massive
satellite galaxies (e.g., Boylan-Kolchin \etal 2010; Busha \etal 2011)
and the detection rate of dark matter substructure via gravitational
lensing (e.g., Vegetti \etal 2010, 2012). In this paper we use subhalo
mass functions and subhalo catalogs from a variety of numerical
simulations that are publicly available, and that have been obtained
using different subhalo finders, to compare subhalo mass functions,
focusing on the massive end.  We confirm the findings by Onions et
al., that the subhalo mass functions are consistent at the 20 percent
level at the low-mass end. At the massive end, though, different
subhalo finders yield subhalo abundances that differ by more than one
order of magnitude! By comparing the simulation results with a new,
semi-analytical model (Jiang \& van den Bosch 2014b), we demonstrate that
subhalo finders that identify subhaloes based purely on density in
configuration space, such as the popular \Subfind and \BDM,
dramatically underpredict the masses, but not the maximum circular
velocities, of massive subhaloes. We also show that the model
predictions are in excellent agreement with the simulation results
when they are analyzed using more advanced subhalo finders that use
phase-space and/or time domain information in the identification of
subhaloes. We discuss a number of implications of our findings, in
particular with regard to the power-law slope of the subhalo mass and
velocity functions.

Throughout we use $m$ and $M$ to refer to the masses of subhaloes and
host haloes, repectively, use $\ln$ and $\log$ to indicate the natural
logarithm and 10-based logarithm, respectively, and express units that
depend on the Hubble constant in terms of $h = H_0/(100\kmsmpc)$.


\section{Description of Data and Model}  
\label{Sec:description}

The main goal of this paper is two-fold. First, we wish to compare
mass and velocity functions of dark matter subhaloes obtained from
numerical simulations using different subhalo finders among each
other, in order to gauge their robustness. Second, we want to compare
these simulation results to predictions from the new, semi-analytical model
of Jiang \& van den Bosch (2014b) in order to assess its reliability.
This section describes the various numerical simulations and subhalo
finders used in our comparison, followed by a brief description of our
semi-analytical model.

\subsection{Numerical Simulations}
\label{sec:sims}

Table~1 lists the various simulations used in this paper. More details
regarding each of these simulations can be found in the references
listed in the final column.  In the case of the Bolshoi and MultiDark
simulations we use publicly available halo catalogs to construct our
own subhalo mass and velocity functions. In the case of the other
simulations, we use published results, mainly in the form of fitting
functions.  As is evident from Table~1, all these simulations adopt
flat $\Lambda$CDM cosmologies, but with slightly different values for
the cosmological parameters.  However, as we will show below, these
mild differences in cosmology have a negligible impact on the subhalo
mass functions, and one can therefore compare results from these
simulations without having to make any corrections for differences in
cosmology.

\subsection{Subhalo Finders}
\label{sec:finders}

The subhalo mass functions presented below have been obtained using
four different subhalo finders: \Subfind, \BDM, \Rockstar and \SURV.
We briefly describe each of these subhalo finders in turn, but refer
the reader to the original papers, referenced below, for more details.
\begin{table*}\label{tab:simdata}
\caption{Numerical Simulations used in this Paper}
\begin{center}
\begin{tabular}{lccccccrrcl}
\hline\hline
 Simulation & $\Omega_{\rmm,0}$ & $\Omega_{\Lambda,0}$ & $\Omega_{\rmb,0}$ & $\sigma_8$ & $n_\rms$ & $h$ & $L_{\rm box}$ & $N_\rmp$ & $m_\rmp$ & Reference \\
  (1) & (2) & (3) & (4) & (5) & (6) & (7) & (8) & (9) & (10) & (11) \\
\hline
Bolshoi      & 0.27 & 0.73 & 0.047 & 0.82 & 0.95 & 0.70 &   $250$ & $2048^3$ & $1.4 \times 10^8$ & Klypin \etal (2011) \\ 
MultiDark    & 0.27 & 0.73 & 0.047 & 0.82 & 0.95 & 0.70 &  $1000$ & $2048^3$ & $8.6 \times 10^9$ & Prada  \etal (2012) \\ 
Millennium I  & 0.25 & 0.75 & 0.045 & 0.90 & 1.0  & 0.73 &   $500$ & $2160^3$ & $8.6 \times 10^8$ & Springel \etal (2005) \\ 
Millennium II & 0.25 & 0.75 & 0.045 & 0.90 & 1.0  & 0.73 &   $100$ & $2160^3$ & $6.9 \times 10^6$ & Boylan-Kolchin \etal (2010) \\ 
Millennium HS & 0.25 & 0.75 & 0.045 & 0.90 & 1.0  & 0.73 &   $100$ &  $900^3$ & $9.5 \times 10^7$ & Angulo \etal (2009) \\ 
Aquarius     & 0.25 & 0.75 & 0.045 & 0.90 & 1.0  & 0.73 & zoom-in &  zoom-in & $>1.7 \times 10^3$ & Springel \etal (2008) \\ 
Phoenix      & 0.25 & 0.75 & 0.045 & 0.90 & 1.0  & 0.73 & zoom-in &  zoom-in & $>6.4 \times 10^5$ & Gao \etal (2012) \\ 
Rhapsody     & 0.25 & 0.75 & 0.040 & 0.80 & 1.0  & 0.73 & zoom-in &  zoom-in & $1.3 \times 10^8$ & Wu \etal (2013) \\ 
GIF2         & 0.30 & 0.70 & 0.040 & 0.90 & 1.0  & 0.70 &   $110$ &  $400^3$ & $1.7 \times 10^9$ & Gao \etal (2004) \\ 
\hline\hline
\end{tabular}
\end{center}
\medskip
\begin{minipage}{\hdsize}
  Parameters of the various numerical simulations discussed in this
  paper.  Columns (2) - (7) list the present-day cosmological density
  parameters for the matter, $\Omega_{\rmm,0}$, the cosmological
  constant, $\Omega_{\Lambda,0}$, and the baryonic matter,
  $\Omega_{\rmb,0}$, the normalization, $\sigma_8$, and spectral
  index, $n_\rms$, of the matter power spectrum, and the Hubble
  parameter, $h = H_0/(100\kmsmpc)$. Columns (8) - (10) list the box
  size of the simulation, $L_{\rm box}$, (in $h^{-1} \Mpc$), the
  number of particles used, $N_\rmp$, and the particle mass, $m_\rmp$
  (in $h^{-1} \Msun$), respectively.  Note that Aquarius, Phoenix and
  Rhapsody are ensembles of high-resolution zoom-in simulations of
  MW-sized (Aquarius) and cluster-sized (Phoenix \& Rhapsody) haloes.
  More details regarding each simulation can be found in the
  references listed in Column~(11).
\end{minipage}
\end{table*}

\subsubsection{Bound-Density-Maximum}
\label{sec:BDM}

The Bound-Density-Maximum (\BDM) algorithm, developed by Klypin \&
Holtzman (1997), identifies both host haloes (also called `distinct'
haloes) and subhaloes (see also Riebe \etal 2013). It locates density
maxima in the particle distribution and uses an iterative scheme to
remove unbound particles.  As described in Appendix~A of Klypin \etal
(2011), if two haloes are (i) separated by less than one virial
radius, (ii) have masses that differ by less than a factor of 1.5, and
(iii) have a relative velocity less than 0.15 of the rms velocity of
dark matter particles inside the haloes, \BDM removes the smaller halo
and keeps only the larger one. As we show below, and as anticipated in
Klypin \etal (2011), this `correction', which was introduced ``to
remove a defect of halo-finding where the same halo is identified more
than once'' results in systematic errors in the subhalo mass function
at the massive end, especially in more massive host haloes (which
assembled more recently).  In this paper we use \BDM catalogs obtained
from both the Bolshoi and MultiDark simulations.
\begin{figure}
\centerline{\psfig{figure=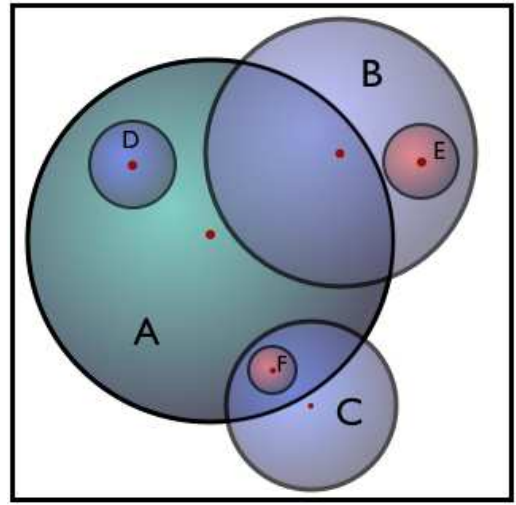,width=0.8\hssize}}
\caption{Illustration of overlapping (spherical) haloes. The small red
  dots indicate the halo centers. Since the centers of B, D and F are
  all located within the extent of A, most subhalo finders will
  consider them subhaloes of A.  Note, though, that B lies partially
  outside of A, so that not all of B's mass is considered part of A,
  at least not in \Rockstar and \BDM. Halo E is a subhalo of B, which
  makes it a second-order subhalo of A. However, since it's center
  falls outside of A, \Rockstar and \BDM do not count it as
  such. Finally, F is a first-order subhalo of C, and will be counted
  as such in the semi-analytical model. However, since its center falls
  inside of A, while that of C doesn't, both \Rockstar and \BDM will
  consider F a first-order subhalo of both A and C. See
  \S\ref{sec:massdef} for a detailed discussion.}
\label{fig:illustration}
\end{figure}
 
\subsubsection{SUBFIND}
\label{sec:Subfind}

The \Subfind algorithm, developed by Springel \etal (2001), is similar
to \BDM in that it identifies substructures within a host halo by
searching for overdense regions using a local density estimate,
obtained by kernel interpolation over the nearest neighbors.  It
identifies substructure candidates as regions bounded by an isodensity
surface that traverses a saddle point of the density field and uses an
iterative unbinding procedure to ensure that these potential
substructures are physically bound. \Subfind has been used extensively
in analyzing the Millennium simulations (Boylan-Kolchin \etal 2009,
2010; Gao \etal 2011), the Aquarius simulations (Springel \etal 2008),
and the Phoenix simulations (Gao \etal 2012). In what follows we use
some of these results (mainly in the form of published fitting
functions) for comparison.

\subsubsection{ROCKSTAR}
\label{sec:Rockstar}

\Rockstar (Robust Overdensity Calculation using K-Space Topologically
Adaptive Refinement) is a phase-space halo finder designed to maximize
halo consistency across time-steps (Behroozi \etal 2013a,b). It uses
adaptive hierarchical refinement of friends-of-friends groups in six
phase-space dimensions and one time dimension, resulting in a very
robust tracking of substructure (see Knebe \etal 2011,
2013). \Rockstar has been used to analyze, among others, the Bolshoi,
MultiDark and Rhapsody simulations.

\subsubsection{SURV}
\label{sec:SURV}

The subhalo finder \SURV was developed by Tormen \etal (1998), and
improved upon by Tormen, Moscardini \& Yoshida (2004) and Giocoli
\etal (2008a).  It identifies subhaloes within the virial radius of a
final host halo by following their progenitors from the time they were
first accreted by the host's main progenitor. Hence, \SURV differs
from the methods discussed above, in that it uses prior information
based on the host halo's merger history to identify its subhaloes. In
particular, subhaloes are identified as those subsets of particles
that belonged to one and the same progenitor halo at its moment of
accretion (i.e., when it first became a subhalo) that are still part
of a self-bound entity within the corresponding tidal radius (see
\S\ref{sec:massdef} below). \SURV is very similar to the Hierarchical
Bound Tracing (\HBT) method recently developed by Han \etal (2012). In
this paper we use the subhalo mass functions obtained by Giocoli \etal
(2008a) using a \SURV analysis of the GIF2 simulation.

\subsection{Analytical Model}
\label{sec:model}

In addition to the simulation results mentioned above, we also include
in our comparison results from the semi-analytical model developed by the
authors, and described in detail in our companion paper (Jiang \& van
den Bosch 2014b; hereafter Paper~I). Starting from halo merger trees,
this model uses a simple semi-analytical description for the {\it average}
subhalo mass loss rate (where the average is taken over all orbital
energies, orbital angular momenta and orbital phases) to evolve the
masses of dark matter subhaloes from the moment of accretion to
$z=0$. It is a modified and improved version of the model developed by
van den Bosch, Tormen \& Giocoli (2005). The main improvement is that
the merger trees are constructed using the algorithm developed by
Parkinson, Cole \& Helly (2008), which, as discussed in Jiang \& van
den Bosch (2014a), is far more reliable than the Somerville \& Kolatt
(1999) method used in van den Bosch \etal (2005). In addition, it uses
an improved model for the subhalo mass loss rate that is calibrated
against numerical simulations, includes scatter in the mass loss rates
that arises from the variance in orbital energies and angular momenta,
and treats the entire hierarchy of substructure, including
sub-subhaloes, sub-sub-subhaloes, etc.  As shown in Paper~I, this
model can accurately reproduce the subhalo mass functions obtained by
Giocoli \etal (2008a) and Wu \etal (2013) using the GIF2 and Rhapsody
simulations, respectively.  One of the main goals of this paper is to
compare this simple and fast semi-analytical model to other simulation
results.

In addition to subhalo masses, the semi-analytical model of Jiang \& van
den Bosch (2014b) also yields the maximum circular velocities, $V_{\rm
  max}$, for both host haloes and subhaloes. For host haloes, $V_{\rm
  max}$ is computed assuming that the density distribution of dark
matter haloes follow a NFW profile (Navarro, Frenk \& White 1997), so
that
\begin{equation}
V_{\rm max} = 0.465\,V_{\rm vir}\,\sqrt{c \over \ln(1+c) - c/(1+c)}\,,
\end{equation}
where $c$ is the halo's concentration parameter, and 
\begin{eqnarray}
\lefteqn{V_{\rm vir} = 159.43 \kms \, \left({M \over 10^{12}\msunh}\right)^{1/3}
\, \left[{H(z) \over H_0}\right]^{1/3}} \nonumber \\
& & \, \left[{\Delta_{\rm vir}(z) \over 178}\right]^{1/6}\,,
\end{eqnarray}
is the virial velocity of a dark matter halo of virial mass $M$ at
redshift $z$. Here $H(z)$ is the Hubble parameters, and $\Delta_{\rm
  vir}(z)$ specifies the average density of a collapsed dark matter
halo in units of the critical density, $\rho_{\rm crit}$, and is well
represented by the fitting function of Bryan \& Norman (1998):
\begin{equation}
\Delta_{\rm vir}(z) = 18 \pi^2 + 82 x - 39 x^2\,,
\end{equation}
where $x = \Omega_\rmm(z) - 1$.  For the cosmologies used in this
paper $\Delta_{\rm vir} \sim 100$. It is well known that the halo
concentration is strongly correlated with the halo's assembly history,
to the extent that more concentrated haloes assemble earlier (e.g.,
Navarro \etal 1997; Wechsler \etal 2002; Giocoli, Tormen \& Sheth
2012; Ludlow \etal 2013).  We use the model of Zhao \etal (2009),
according to which
\begin{equation}
c(M,t) = 4.0 \, \left[ 1+ \left({t \over 3.75 \, t_{0.04}}\right)^{8.4}
\right]^{1/8}\,.
\end{equation}
Here $t_{0.04}$ is the proper time at which the host halo's main
progenitor gained 4\% of its mass at proper time $t$, which we extract
from the halo's merger tree (see Paper~I for details). For subhaloes,
we compute $V_{\rm max}$ using
\begin{equation}\label{Vmaxfit}
V_{\rm max} = 2^{\mu} \, V_{\rm acc} \, {(m/m_{\rm acc})^{\eta} \over
 (1 + m/m_{\rm acc})^{\mu}}\,,
\end{equation}
where $m_{\rm acc}$ and $V_{\rm acc}$ are the subhalo mass and maximum
circular velocity at the time of accretion, and $(\eta,\mu) =
(0.44,0.60)$. As described in detail in Paper~I, this relation between
$V_{\rm max}/V_{\rm acc}$ and $m/m_{\rm acc}$ is obtained by fitting
data from the Rhapsody simulations (Wu \etal 2013), and, as we
demonstrate below, also adequately describes the data from the Bolshoi
and MultiDark simulations. As shown in Paper~I, implementing this
`recipe' for $V_{\rm max}$ yields subhalo velocity functions, $\rmd
N/\rmd\log(V_{\rm max}/V_{\rm vir})$, where $V_{\rm vir}$ is the
virial velocity of the host halo, in excellent agreement with the
simulation results of Wu \etal (2013).

\subsection{Definition of (Sub)Halo Mass}
\label{sec:massdef}

A recurring source of contention in this paper is the exact definition
of (sub)halo mass. Therefore, as a preamble, we now discuss (some of)
the various definitions that are used here and in the literature.

Throughout this study we focus on results at $z=0$, and we define all
{\it host} haloes of mass $M$ as spheres with an average density
$\rho_\rmh \equiv 3 M/4 \pi r^3_{\rm vir} = \Delta_{\rm vir} \rho_{\rm
  crit}$, where $r_{\rm vir}$ is the virial radius.  We follow the
nomenclature of Paper~I and refer to subhaloes that were accreted
directly onto the main progenitor of the host halo as first-order
subhaloes. Similarly, haloes that accrete directly onto the
progenitors of first-order subhaloes give rise to second-order
subhaloes (or sub-subhaloes), and the same logic is used to define
progenitors and subhaloes of third order and higher.

In the case of our semi-analytical model, we define subhalo masses as the
fraction of the virial mass the progenitor had at accretion that has
not been stripped off (as described by our mass-loss model). Hence, in
our model subhalo mass always decreases with time, and we ignore the
possibility that two subhaloes may merge. Also, the model treats all
masses as `inclusive', including in each (sub)halo mass the mass of
all its substructure.  This implies that a single dark matter particle
can be part of multiple sub-structures (of different order). Note
that, in this case, the fraction of host halo mass that is locked up
in substructure is obtained by integrating the SHMF of first-order
subhaloes only.
\begin{figure}
\centerline{\psfig{figure=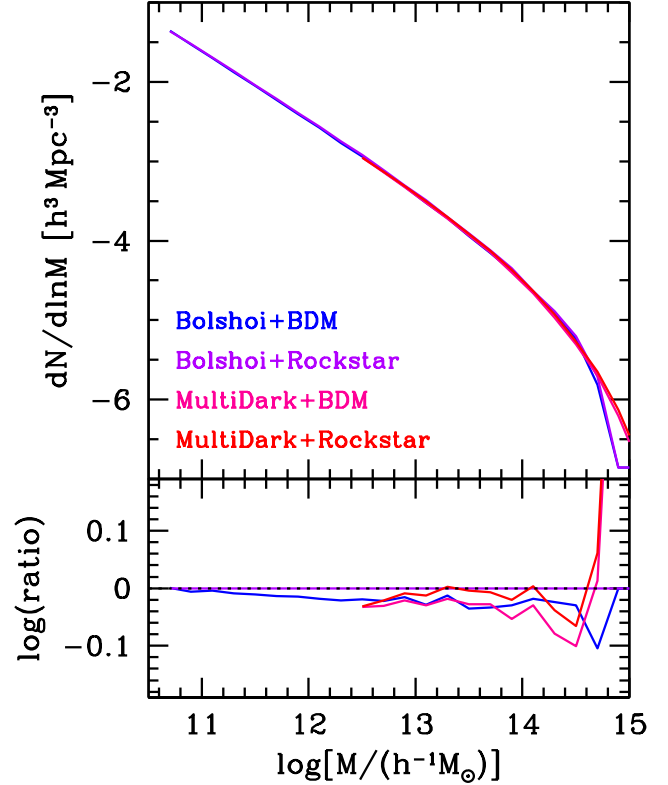,width=\hssize}}
\caption{Mass functions of dark matter host haloes. Results are shown
  for both the Bolshoi and MultiDark simulations analyzed using both
  \BDM and \Rockstar (as indicated). The lower panel plots the
  logarithm of the ratio of these mass functions with respect to that
  of Bolshoi$+$\Rockstar, and shows that the various mass functions
  are in excellent agreement. There are some differences between
  MultiDark and Bolshoi at the massive end due to sample variance, but
  overall the \Rockstar and \BDM results agree at the 10 to 20 percent
  level.}
\label{fig:dNdlnM}
\end{figure}

The subhalo finders \Rockstar and \BDM use the same definitions of
subhalo mass. After the finder has identified the set, $\calS$, of
particles that belong to a self-bound substructure, the center of the
subhalo is identified as the location of its most bound particle (here
the potential is computed using solely the particles in $\calS$).  The
mass of the subhalo is defined as the sum over all particles in
$\calS$ that fall within a radius $r_{\rm vir}$ of this center, where
$r_{\rm vir}$ is defined above. Hence, the subhalo mass can be less
than the mass of all particles in $\calS$.  Both \Rockstar and \BDM
are similar to our semi-analytical model in that their (sub)halo masses are
always `inclusive'. However, some non-trivial issues can arise when
(sub)haloes overlap. As an example, consider the situation illustrated
in Fig.~\ref{fig:illustration}.  Here B is a subhalo of A, since it's
center is located within the virial radius of A. However, only the
mass of B that is located within that same radius is counted towards
the mass of A. Hence, in this case, where A and B have comparable
mass, the mass ratio, $m/M$, of subhalo mass to host halo mass can
exceed $0.5$. Note that this is not possible in the case of our
semi-analytical model. As we will see, this results in subtle differences
at the massive end of the subhalo mass functions.  Note also that
although halo E in Fig.~\ref{fig:illustration} is a subhalo of B, it
is not counted as a sub-subhalo of A in the \Rockstar and \BDM halo
catalogs because its center does not fall within the virial radius of
A.  This is despite the fact that B is considered a subhalo of A. Even
more tricky is the issue of halo F. Before C started to overlap with
A, F was an unambiguous subhalo of C. Since the center of C hasn't yet
entered the virial radius of A, our semi-analytical model still considers F
to be a first-order subhalo of C, unassociated with A. However, since
F falls inside of A, most subhalo finders will count F as a
first-order subhalo of {\it both} C and A. These non-trivial
differences in assigning masses and orders to subhaloes can be
responsible for subtle differences in the mass functions of
(higher-order) subhaloes.

In the case of \SURV, we follow the definition of Giocoli \etal (2008a)
in which the mass of a subhalo is simply defined as the mass of the
subset of particles that belonged to the virial mass of a progenitor
halo at its moment of accretion (i.e., when it first became a subhalo)
that are still part of a self-bound entity within the corresponding
tidal radius. 
As for \Rockstar and \BDM, these (sub)halo masses are `inclusive'.
\begin{figure*}
\centerline{\psfig{figure=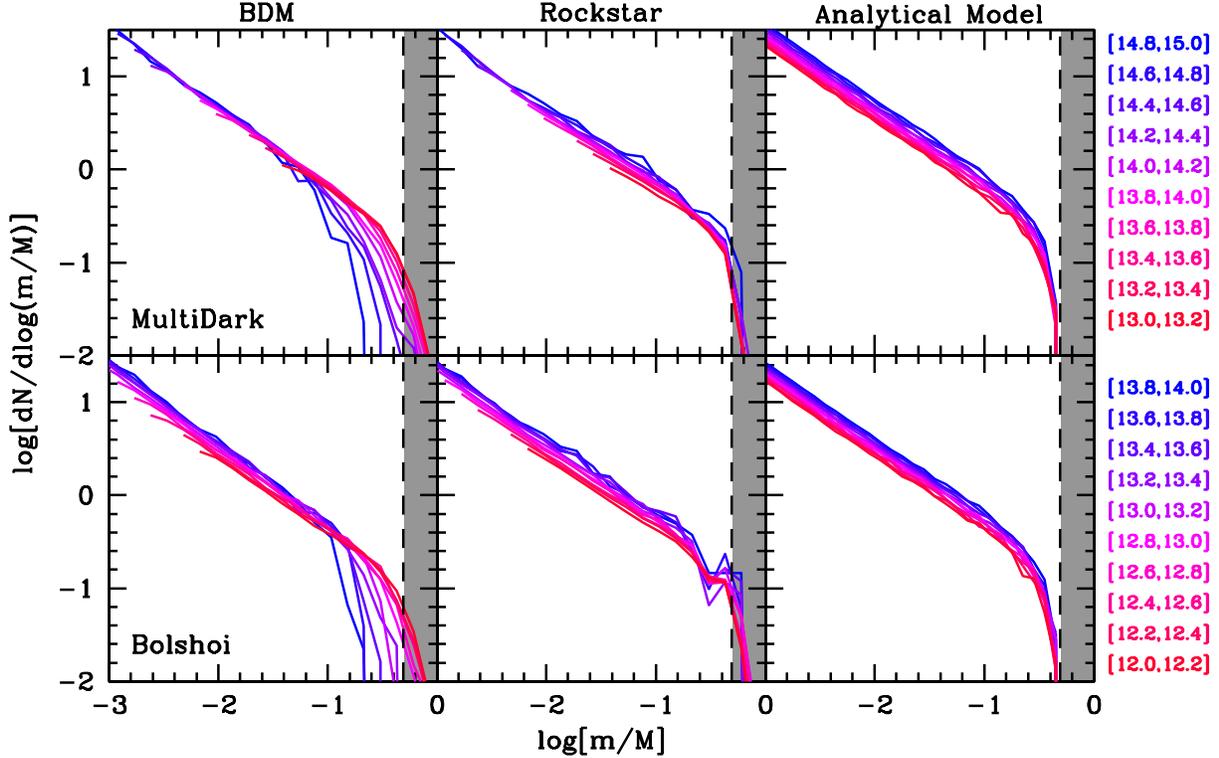,width=0.9\hdsize}}
\caption{Subhalo mass functions, $\rmd N/\rmd\log(m/M)$, for dark
  matter host haloes in different mass bins (different colors), as
  function of the subhalo mass, $m$, normalized to the host halo mass,
  $M$. The values in square brackets at the right side indicate the
  range in host halo mass $\log[M/(h^{-1}\Msun)]$. Panels in
  left and middle columns show SHMFs obtained using the subhalo
  finders \BDM and \Subfind applied to the MultiDark (upper panels)
  and Bolshoi (lower panels) simulations. Results are only shown for
  subhaloes that contain at least 50 particles, which explains why
  less massive host haloes have their SHMF restricted to a more
  limited dynamic range in $m/M$. For comparison, the SHMFs shown in
  the right-hand panel have been obtained using our semi-analytical model,
  averaging over 10,000 host haloes as described in the text. The gray
  regions at the right side of each panel indicate where $m/M > 0.5$,
  which is a forbidden region for our semi-analytical model, but not for the
  simulations (see discussion in \S\ref{sec:massdef}).}
\label{fig:SHMF}
\end{figure*}

Finally, in the case of \Subfind, the mass of a subhalo is defined as
the mass of all self-bound particles located within the isodensity
surface that bounds the object.  \Subfind masses differ from all other
masses discussed above in that they are `exclusive': the mass of a
sub-subhalo is not included in the mass of its `hosting'
subhalo. Consequently, integrating $\rmd N/\rmd(m/M)$ for first-order
subhaloes does {\it not} yield the mass fraction in substructure,
which instead requires integrating the SHMF of {\it all} orders of
subhaloes.  We emphasize, though, that the difference between
`inclusive' and `exclusive' masses is small, and does not have a
significant impact on the SHMFs. This is easy to understand; subhaloes
typically account for only about $\sim 10$ percent of the mass of a
host halo. Similarly, sub-subhaloes only account for $\lta 10$ percent
of the mass of a subhalo, etc. Hence, the difference between inclusive
and exclusive can cause differences of up to $\sim 10$ percent in the
masses of individual subhaloes, which does not have a significant
impact on $\rmd N/\rmd\log(m/M)$, at least not compared to the differences
we are concerned with in this paper (see also Onions \etal 2012).
\begin{figure*}
\centerline{\psfig{figure=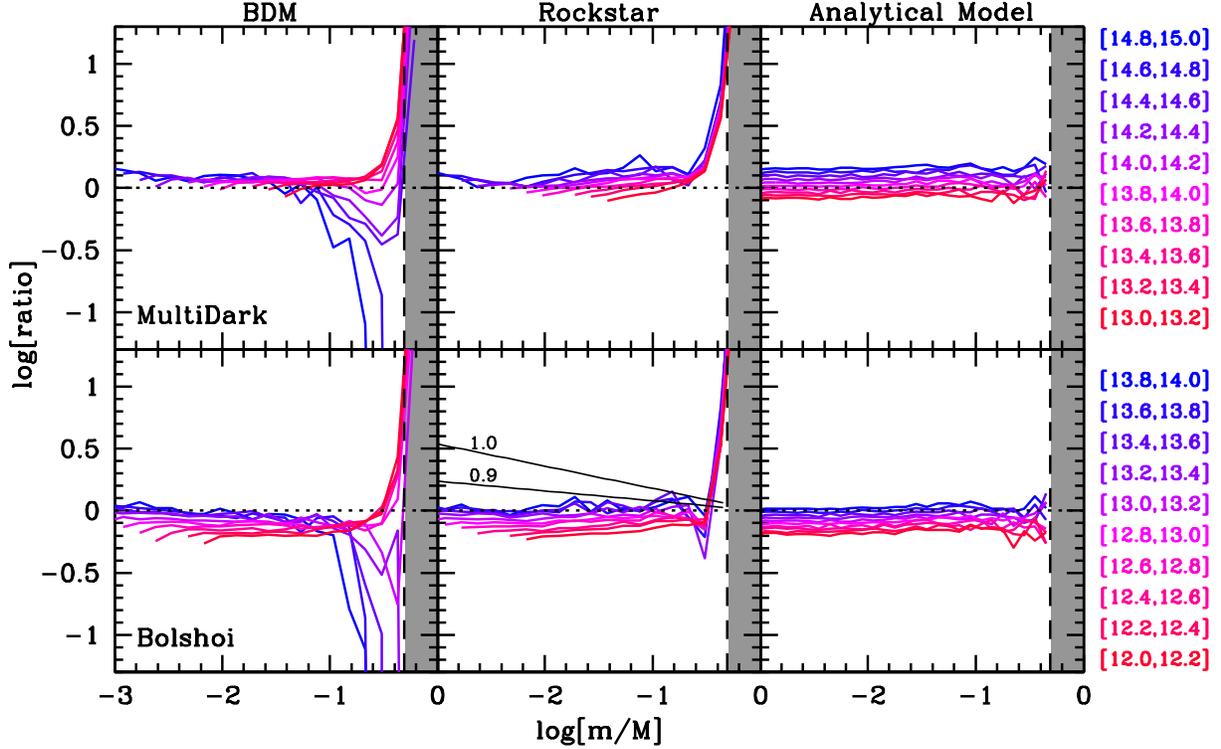,width=0.9\hdsize}}
\caption{Same as Fig.~\ref{fig:SHMF}, except that here we show the
  logarithm of the ratio between the SHMF, $\rmd N/\rmd\log(m/M)$, and
  the fiducial SHMF given by Eq.~(\ref{fiducial}) with slope $\alpha =
  0.82$ and normalization $A_M = 0.09$. Note the good agreement for
  $\log[m/M] \lta -1$, and the large differences at the massive
  end. The black, solid lines in the lower, middle panel indicate the
  residuals that would arise if the SHMF has the form of
  Eq.~(\ref{fiducial}), but with $\alpha = 0.9$ and $1.0$, as labeled.
  This shows that the power-law slopes of the SHMFs are clearly less
  steep than $0.9$, contrary to numerous claims in the literature (see
  text for a detailed discussion).}
\label{fig:diffSHMF}
\end{figure*}


\section{Results}
\label{sec:res}

We start our investigation by comparing the $z=0$ subhalo mass
functions obtained from our semi-analytical model with those obtained from
the Bolshoi and MultiDark simulations using both \BDM and \Rockstar.
The Bolshoi halo catalogs are publicly available at the Bolshoi
website\footnote{http://hipacc.ucsc.edu/Bolshoi/MergerTrees.html},
which also lists the \Rockstar catalogs for the MultiDark simulation.
The \BDM catalog for the MultiDark simulation is downloaded from the
MultiDark website\footnote{http://www.multidark.org/MultiDark/}.  For
each host halo, the catalogs contain the present-day virial mass,
$M$, and maximum circular velocity, $V_{\rm max}$.  In the
case of subhaloes, the output contains the present-day mass of the
subhalo, $m$, its maximum circular velocity, $V_{\rm max}$, as well as
the virial mass, $m_{\rm acc}$, and maximum circular velocity, $V_{\rm
  acc}$, of the subhalo at the time of accretion.

Before comparing subhalo statistics, it is useful to check that the
different simulations and halo finders yield consistent abundances of
{\it host} haloes.  Fig.~\ref{fig:dNdlnM} compares the halo mass
functions of host haloes in MultiDark and Bolshoi, as obtained with
both \Rockstar and \BDM. As is evident, the agreement is excellent.
The Bolshoi results deviate from the MultiDark results at the massive
end, but this is a manifestation of sample variance arising from the
limited box sizes. Overall, the abundances obtained using \Rockstar
and \BDM agree with each other at the 10 to 20 percent level, in good
agreement with the results of Knebe \etal (2011).

\subsection{Subhalo Mass Functions: Bolshoi \& MultiDark}
\label{sec:SHMF1}

We now turn our attention to subhaloes.  Using all haloes and
subhaloes with at least 50 dark matter particles, we compute the
SHMFs, $\rmd N/\rmd\log(m/M)$, for 10 different bins in host halo
mass, each with a bin width of 0.2 dex. In the case of MultiDark we
use logarithmic mass bins $\log[M/(h^{-1}\Msun)] \in [13 + 0.2\,(n-1),
  13 + 0.2\,n]$ with $n=(1,2,...,10)$, while for the smaller but
higher-resolution Bolshoi simulation, we adopt $[12 + 0.2\,(n-1), 12 +
  0.2\,n]$. When counting subhaloes, we include subhaloes of all
orders (i.e., subhaloes, sub-subhaloes, etc.).  The resulting SHMFs
are shown in Fig.~\ref{fig:SHMF}: upper and lower panels correspond to
Multidark and Bolshoi, respectively, while panels in the left and
middle columns show the results obtained using the \BDM and \Rockstar
catalogs, respectively.  Different colors correspond to different bins
in host halo mass, as indicated. For comparison, the right-hand panels
show the SHMFs obtained using our semi-analytical model averaged over
10,000 host haloes with masses $\log[M/(h^{-1}\Msun)] = 13.1 + 0.2\,n$
(MultiDark) and $12.1 + 0.2\,n$ (Bolshoi). Here we have adopted the
same cosmology as for the simulations (see Table~1), and again the
SHMFs include subhaloes of all orders. The gray area at the right side
of each panel marks the region where $m/M > 0.5$; no subhalo in our
semi-analytical model can have a mass this large. However, as is evident,
some of the \Rockstar and \BDM SHMFs do extent into this regime. As
discussed in \S\ref{sec:massdef}, this as a consequence of their
treatment of sub-haloes that are not entirely located within their
host halo.

In order to facilitate a more detailed comparison,
Fig.~\ref{fig:diffSHMF} shows the ratios of all these SHMFs with
respect to a fiducial SHMF given by
\begin{equation}\label{fiducial}
{\rmd N \over \rmd\log(m/M)} = A_M \, \left({m \over M}\right)^{-\alpha} \,
\exp\left[-50\,(m/M)^4\right]\,.
\end{equation}
with slope $\alpha = 0.82$ and normalization $A_M = 0.09$. As shown in
Paper~I, this functional form accurately fits the SHMFs predicted by
our semi-analytical model, while the normalization parameter $A_M$ is a
function of halo mass, redshift and cosmology. This is also evident
from the right-hand panels of Figs.~\ref{fig:SHMF}
and~\ref{fig:diffSHMF}, which show that the functional form of the
model SHMF is invariant, but that more massive haloes have more
substructure (i.e., a larger normalization $A_M$). This arises because
(i) the {\it unevolved} subhalo mass function (i.e., the mass function
of the subhaloes at accretion) is independent of host halo mass, and
(ii) more massive haloes assemble later.  Consequently, their
subhaloes have, on average, been exposed to mass stripping for a
shorter period of time (see also van den Bosch \etal 2005 and
Paper~I).

Overall, the simulation results are in good agreement with each other
and with the model predictions. This is especially true at the low
mass end ($m \lta 0.1 M$) where the agreement is at the level of $\sim
20$ percent ($\Delta\log[\rmd N/\rmd\log(m/M)] \lta 0.1$dex),
confirming the earlier findings of Onions \etal (2012). Particularly
reassuring is the fact that the simulation results reveal the same
host halo mass dependence as predicted by our model (see also Gao
\etal 2004 and Shaw \etal 2006). Note, though, that for some unknown
reason, the SHMFs obtained from the MultiDark+\BDM analysis do not
reveal any significant dependence on host halo mass at the low mass
end.
\begin{table}\label{tab:Nhost}
\caption{Number of host haloes}
\begin{center}
\begin{tabular}{llrrr}
\hline\hline
 Simulation & Algorithm & $[12,12.5]$ & $[13,13.5]$ & $[14,14.5]$ \\
\hline
Bolshoi     & \Rockstar &      37,498 &       4,281 &       283 \\
Bolshoi     & \BDM      &      35,889 &       4,046 &       273 \\
MultiDark   & \Rockstar &        --   &     267,745 &     17,699 \\
MultiDark   & \BDM      &        --   &     256,399 &     16,342 \\
MSII        & \Subfind  &       2,039 &        --   &        --  \\
GIF2        & \SURV     &       3,349 &         461 &         35 \\
\hline\hline
\end{tabular}
\end{center}
\medskip
\begin{minipage}{\hssize}
  Number of host haloes used for the SHMFs shown in
  Fig.~\ref{fig:comparison}.  Columns (1) and (2) indicate the
  simulation and subhalo finder, while columns (3) - (5) list the
  number of host haloes in the mass bin indicated at the top, where
  the values in square brackets mark the range in
  $\log[M/(h^{-1}\Msun)]$.
\end{minipage}
\end{table}

Another important result is that all SHMFs agree, to good accuracy, on
the slope at the low-mass end, $\alpha$, which is equal to 0.82 in the
case of our semi-analytical model (see Eq.~[\ref{fiducial}] and
Paper~I). This value for $\alpha$ falls roughly midway of the values
reported by studies based on numerical simulations, which span the
entire range from $0.7$ to $1.1$ (Moore \etal 1999; Ghigna \etal 1999;
Helmi \etal 2002; De Lucia \etal 2004; Gao \etal 2004; Shaw \etal
2006; Diemand, Kuhlen \& Madau 2007; Angulo \etal 2009; Giocoli \etal
2010). There are a number of reasons for this large spread in measured
slopes. Foremost, earlier studies lacked sufficient numerical
resolution and/or statistics, and where only able to fit the slope of
the SHMF over a limited range in mass. In addition, many of these
studies fitted the SHMF with a single power-law, without an
exponential tail, which tends to bias the slope high. And finally,
some of the discrepancy arises from the use of different subhalo
finders. In particular, in a recent study of Milky Way sized haloes
extracted from the Millennium II (Boylan-Kolchin \etal 2009) and
Aquarius (Springel \etal 2008) simulations, Boylan-Kolchin \etal
(2010) used \Subfind to infer a slope $\alpha = 0.94$. Gao \etal
(2012), using \Subfind to analyze nine high-resolution simulations of
cluster-sized haloes that are part of the `Phoenix Project', inferred
an even steeper slope of $\alpha = 0.98$, very close to the critical
value of unity for which each logarithmic mass bin contributes equally
to the total mass in substructure. Such a steep slope is not only
inconsistent with our model predictions, but also with the \BDM and
\Rockstar results shown, all of which suggest that $0.78 \lta \alpha
\lta 0.84$. To emphasize this level of inconsistency, the thin, solid
lines in the middle, lower panel of Fig.~\ref{fig:diffSHMF} correspond
to Eq.~(\ref{fiducial}) but with $\alpha = 0.9$ and $1.0$, as
indicated. We believe that this discrepancy is due to issues with
\Subfind, which, as we demonstrate below, yields SHMFs that strongly
deviate from most other subhalo finders, especially at the massive
end.
\begin{figure*}
\centerline{\psfig{figure=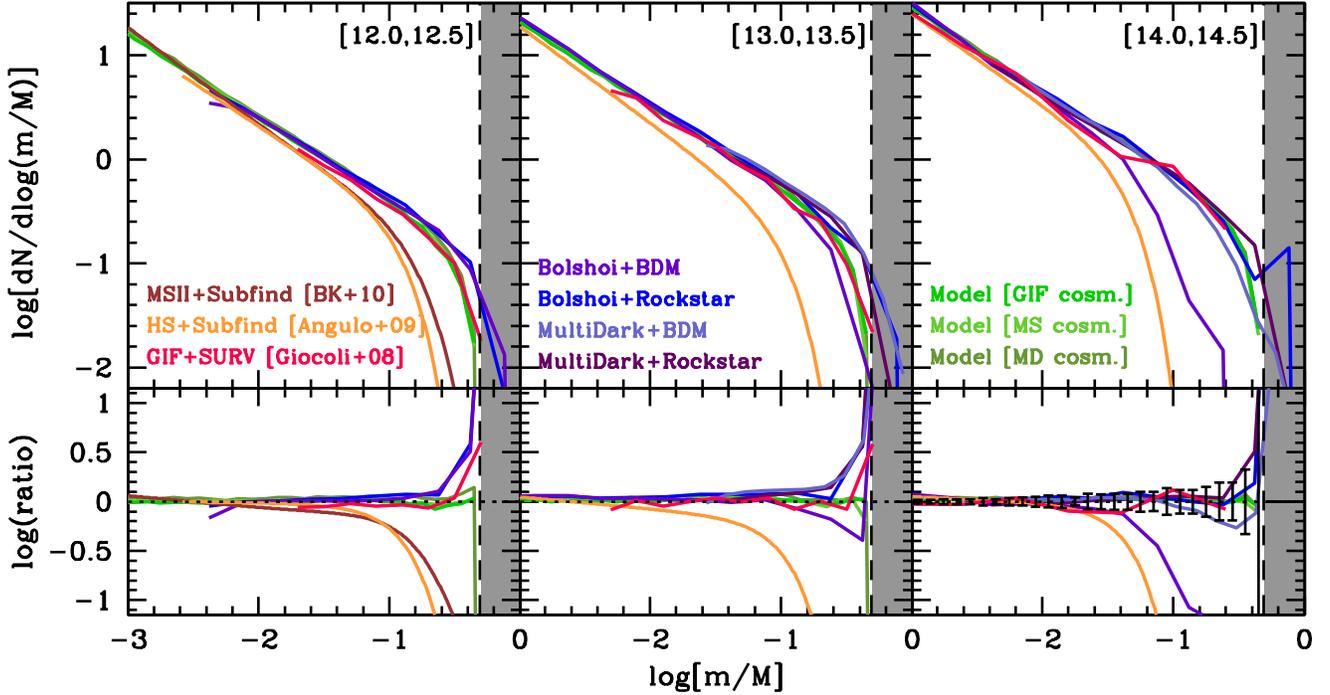,width=0.98\hdsize}}
\caption{{\it Upper panels:} Subhalo mass functions, $\rmd
  N/\rmd\log(m/M)$, as function of the subhalo mass, $m$, normalized
  to the host halo mass, $M$. Results are shown for three bins in host
  halo mass, $\log[M/(h^{-1}\Msun)]$, as indicated by the values in
  square brackets in the upper-right corner of each column.  Different
  curves correspond to different model predictions (green-colored
  curves) or simulation results (curves of other colors), as
  indicated. See text for a detailed discussion {\it Lower panels:}
  Same as upper panels, except that here we show the logarithm of the
  ratio between the SHMF, $\rmd N/\rmd\log(m/M)$, and the fiducial
  SHMF given by Eq.~(\ref{fiducial}) with $A_M = 0.060$ (left panel)
  $0.076$ (middle panel), and $0.102$ (right panel).}
\label{fig:comparison}
\end{figure*}
  
Although Figs.~\ref{fig:SHMF} and~\ref{fig:diffSHMF} indicate a good
level of agreement among simulations and model for subhaloes with $m/M
\lta 0.1$, the situation is less cheerful at the massive end, not
probed by the comparison study of Onions \etal (2012). Whereas
\Rockstar yields SHMFs that show very little dependence on host halo
mass at the exponential tail of $\rmd N/\rmd\log(m/M)$, in excellent
agreement with the predictions of our semi-analytical model, the \BDM
results are very different. In particular, \BDM yields SHMFs for which
the abundance of massive subhaloes declines strongly with increasing
host halo mass. In host haloes with $M \gta 3 \times 10^{14} h^{-1}
\Msun$, the \BDM and \Rockstar abundances of subhaloes with $m \gta
0.3 M$ differ by more than a factor three!  We believe that this is an
artifact of the `correction' introduced in \BDM to prevent the same
halo from being identified more than once (see \S\ref{sec:BDM}), and
we conclude that SHMFs obtained with \BDM cannot be trusted for $m
\gta 0.1\,M$.

\subsection{Subhalo Mass Functions: Subfind \& Surv}
\label{sec:SHMF2}

Fig.~\ref{fig:comparison} presents another comparison of
SHMFs. Results are shown for three different mass bins:
$\log[M/(h^{-1}\Msun)] \in [12.0,12.5]$ (left-hand panels),
$[13.0,13.5]$ (middle panels) and $[14.0,14.5]$ (right-hand
panels). The blue/purple colored lines are for the Bolshoi and
MultiDark simulations analyzed with \BDM and \Rockstar, as
indicated. These SHMFs have been obtained using the same halo catalogs
and methodology as in Fig.~\ref{fig:SHMF}. The red curves are the
SHMFs obtained by Giocoli \etal (2008a) using the subhalo finder \SURV
applied to the GIF2 simulations. The brown-red curve labeled
'MSII+Subfind' is the (fit to the) SHMF obtained by Boylan-Kolchin
\etal (2010) using a \Subfind analysis of the Millennium Simulation II
(MSII). Since this analysis only focused on Milky-Way sized dark
matter haloes, no SHMFs are available for the $[13.0,13.5]$ and
$[14.0,14.5]$ mass bins. The orange lines are the SHMFs obtained by
Angulo \etal (2009) using a \Subfind analysis of both the Millennium I
and Millennium HS (HS) simulations. Note that Angulo \etal (2009)
define host haloes as spheres with a radius of $r_{200c}$, inside of
which the mean density is 200 times the critical density. Hence, we
need to convert their SHMFs taking account of two effects; first of
all, we need to convert their host halo masses to our definition of
virial mass, $M$, which we do assuming that dark matter haloes follow
an NFW density distribution (Navarro, Frenk \& White 1997) with a
concentration-mass relation given by Macc\`io, Dutton \& van den Bosch
(2008). Secondly, since the volume enclosed by $r_{200c}$ is much
smaller than that enclosed by $r_{\rm vir}$, we need to correct the
subhalo abundances as well. We do so by simply multiplying the
abundances of Angulo \etal by a factor $(r_{\rm
  vir}/r_{200c})^3$. Although this make the oversimplified assumption
that subhaloes are homogeneously distributed within their host halo,
the resulting SHMF for the host mass bin $\log[M/(h^{-1}\Msun)] \in
[12.0,12.5]$ is in excellent agreement with that of Boylan-Kolchin
\etal (2010), which is also based on a \Subfind analysis of the
Millennium simulations\footnote{If, instead, we assume that subhaloes
  follow the dark matter distribution, the correction factor
  is $M/M_{200c}$, which results in a $\rmd N/\rmd\log(m/M)$ that is
  $\sim 0.3$dex lower than all other SHMFs shown at the low mass
  end.}. Furthermore, the resulting SHMFs are in good agreement with
the other SHMFs at the low mass end, indicating that this correction
factor is appropriate.  Finally, the three green lines are the
semi-analytical model predictions for the cosmologies used for the GIF
simulations, the Millennium simulations, and the Bolshoi/MultiDark
simulations. As shown in Table~1, these cosmologies differ slightly;
however, as is evident from a comparison of the three different green
curves, which lie virtually on top of each other, this cosmology
dependence has a completely negligible impact on the SHMFs.

The semi-analytical model predictions are obtained averaging the SHMFs of
25,000 host haloes with masses $M = 10^{12.25} h^{-1}\Msun$ (left-hand
panels), $10^{13.25} h^{-1}\Msun$ (middle panels), and $10^{14.25}
h^{-1}\Msun$ (right-hand panels). The numbers of host haloes used in
each of the other SHMFs are listed in Table~2, except for those of
Angulo \etal (2009), for which this information is not available. They
range from a meager 35, in the case of the \SURV SHMF for the
$[14.0,14.5]$ mass bin to a staggering $\sim 250,000$ for the
MultiDark $[13.0,13.5]$ mass bin. In order to gauge how small number
statistics impacts these SHMFs, we use our semi-analytical model to compute
the average SHMFs of 280 host haloes, representative of the number of
host haloes in the Bolshoi simulation in the $[14.0,14.5]$ mass bin,
using 50 realizations (i.e., using 50 samples of 280 host haloes). The
variance among those 50 average SHMFs is indicated by the errorbars in
the lower-right panel.
\begin{figure*}
\centerline{\psfig{figure=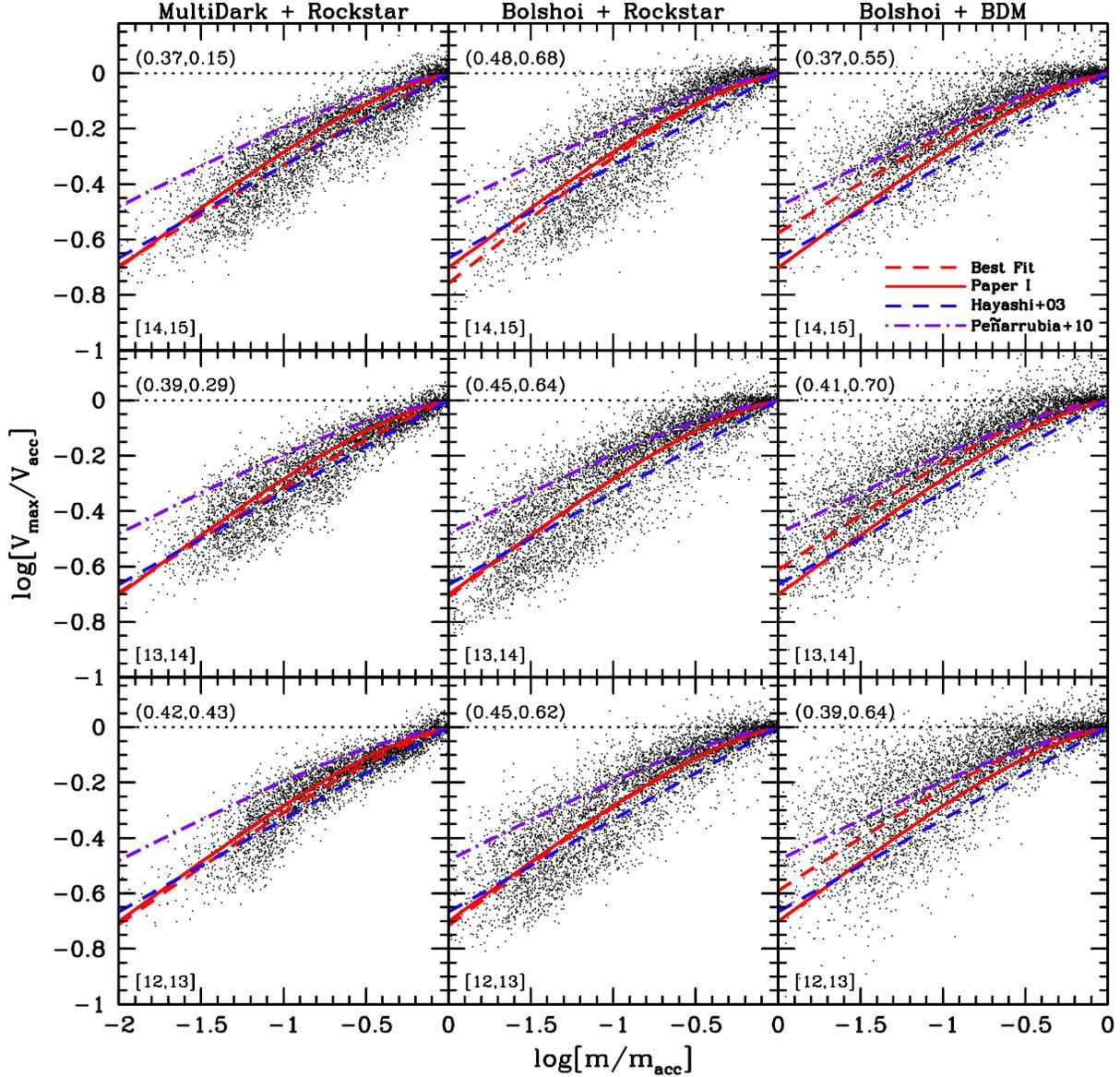,width=0.9\hdsize}}
\caption{The ratio $V_{\rm max}/V_{\rm acc}$ as function of $m/m_{\rm
    acc}$ for dark matter subhaloes obtained from
  MultiDark$+$\Rockstar (left column), Bolshoi$+$\Rockstar (middle
  column) and Bolshoi$+$\BDM (right column).  Different rows
  correspond to different bins in host halo mass, with the range in
  $\log[M/(h^{-1}\Msun)]$ indicated by the values in square brackets
  in the lower left-hand corner of each panel. In order to better
  sample the dependence on $m/m_{\rm acc}$ we plot no more than 150
  subhaloes per 0.05 dex in $\log(m/m_{\rm acc})$.  As shown in
  Paper~I, the $V_{\rm max}/V_{\rm acc}$ - $m/m_{\rm acc}$ relation is
  well described by Eq.~(\ref{Vmaxfit}) with $(\eta,\mu) =
  (0.44,0.60)$, which is indicated by the solid, red line.  The
  dashed, red lines are the best-fit relations of the
  form~(\ref{Vmaxfit}), fit separately to the data in each panel. The
  corresponding best-fit values for $\eta$ and $\mu$ are indicated in
  parenthesis in the upper left-hand corner of each panel. The blue
  dashed and purple dot-dashed curves are the best-fit results of
  Hayashi \etal (2003) and Pe\~narrubia \etal (2010), and are shown
  for comparison.}
\label{fig:massloss}
\end{figure*}

Upon inspection of Fig.~\ref{fig:comparison} it is clear that all
SHMFs agree at the low mass end ($m \lta M/100$) to an accuracy of
$\sim 20$ percent. This once more confirms the findings of Onions
\etal (2012), who compared many more subhalo finders using a
high-resolution simulation of a single host halo from the Aquarius
project. However, Fig.~\ref{fig:comparison} also shows that the
\Subfind SHMFs have a steeper slope at the low mass end than all other
SHMFs (this is particularly apparent for the $[13.0,13.5]$ mass bin),
and that the agreement at the massive end of the SHMF is much weaker,
especially for $m/M \gta 0.1$. In particular;
\begin{itemize}

\item The \Rockstar SHMFs typically predict the largest abundances of
  massive subhaloes, even larger than those predicted by our
  semi-analytical model by about a factor of three for $m/M = 0.4$. As
  discussed above, the fact that the \Rockstar SHMFs venture into the
  regime with $m/M > 0.5$ indicates that this discrepancy with respect
  to the semi-analytical model can largely be explained as a consequence of
  how sub-haloes are treated that are not entirely located within
  their host halo (see \S\ref{sec:massdef}).

\item The \BDM SHMFs are generally in good agreement with those
  obtained with \Rockstar. An exception is the Bolshoi $+$ \BDM SHMF
  for host haloes in the $[14.0,14.5]$ mass bin, which dramatically
  underpredicts the abundance of massive subhaloes with respect to
  \Rockstar, \SURV and our semi-analytical model. As is evident from the
  errorbars in the lower-right panel, this is not a manifestation of
  sample variance. Rather, as already discussed in connection to
  Figs.~\ref{fig:SHMF}-\ref{fig:diffSHMF}, this is most likely due to
  the issues with massive, overlapping haloes discussed in
  \S\ref{sec:BDM}.

\item The GIF $+$ \SURV results of Giocoli \etal (2008a) are in
  excellent agreement with our semi-analytical model (see also
  Paper~I). This is particularly reassuring given that our model uses
  subhalo mass loss rates that have been calibrated against the same
  data set.

\item The SHMFs obtained using \Subfind dramatically underpredict the
  abundances of massive subhaloes. For cluster-sized host haloes the
  discrepancy with the semi-analytical model and with the \SURV and
  \Rockstar results exceeds one order of magnitude for $m/M = 0.1$!

\end{itemize}


\subsection{Evolution of Structural Parameters}
\label{sec:struc}

In addition to the SHMFs, $\rmd N/\rmd\log(m/M)$, we also consider the
subhalo velocity functions, $\rmd N/\rmd\log(V_{\rm max}/V_{\rm
  vir})$, where $V_{\rm vir}$ is the virial velocity of the host halo.
As discussed in \S\ref{sec:model} above, in our semi-analytical model
$V_{\rm max}$ for the subhaloes is computed using a relation between
$V_{\rm max}/V_{\rm acc}$ and $m/m_{\rm acc}$ that is calibrated
against high-resolution simulations of cluster-sized dark matter
haloes from the Rhapsody project (see Paper~I).
Fig.~\ref{fig:massloss} plots $V_{\rm max}/V_{\rm acc}$ versus
$m/m_{\rm acc}$ for subhaloes in the MultiDark and Bolshoi
simulations, obtained using either \BDM or \Rockstar. Results are
shown for three different bins in host-halo mass, as indicated in the
lower left-hand corner of each panel. In order to better delineate the
trends, we plot no more than 150 subhaloes per bin of 0.05 dex in
$\log(m/m_{\rm acc})$. The red, solid line in each panel corresponds
to Eq.~\ref{Vmaxfit} with $(\eta,\mu) = (0.44,0.60)$ which is the
relation that best fits the Rhapsody data and which we use in our
semi-analytical model to compute $V_{\rm max}$. The relation between
$V_{\rm max}/V_{\rm acc}$ and $m/m_{\rm acc}$ has also been studied by
Hayashi \etal (2003) and Pe\~narrubia \etal (2008, 2010) using
high-resolution, idealized $N$-body simulations of individual
subhaloes orbiting in a static, spherical NFW host halo. Hayashi \etal
found that $V_{\rm max}/V_{\rm acc} \propto (m/m_{\rm acc})^{1/3}$,
which is indicated by a dashed, blue line in Fig.~\ref{fig:massloss},
and which is in reasonable agreement with the results obtained in
Paper~I. Pe\~narrubia \etal (2010) fitted their
results using the functional form of Eq.~(\ref{Vmaxfit}). Their
best-fit has $(\eta,\mu) = (0.30,0.40)$ and is indicated as a purple
dot-dashed curve in Fig.~\ref{fig:massloss}. Finally, we use the
Levenberg-Marquardt method to fit the data in each panel with
Eq.~(\ref{Vmaxfit}), treating $\eta$ and $\mu$ as free parameters. The
best-fit values for $\eta$ and $\mu$ are indicated in parenthesis in
the upper left-hand corner of each panel, while the red, dashed line
shows the corresponding best-fit relation.
\begin{figure*}
\centerline{\psfig{figure=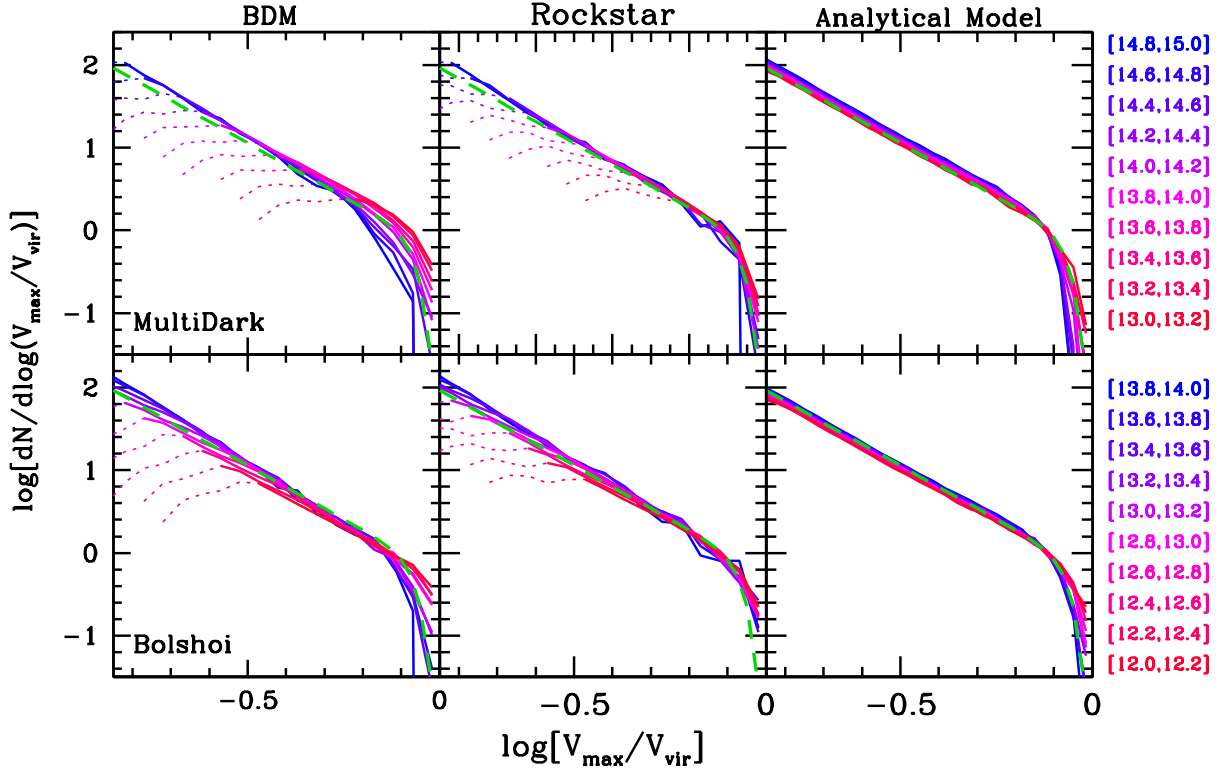,width=0.9\hdsize}}
\caption{Same as Fig.~\ref{fig:SHMF} except here we show the subhalo
  velocity functions $\rmd N/\rmd\log(V_{\rm max}/V_{\rm vir})$. Solid
  lines are the results for all subhaloes with at least 250 particles,
  while the dotted parts of the curves show the extensions one obtains
  when including subhaloes down to a limit of 50 particles per
  subhalo.  The green, dashed curve in each panel corresponds to
  Eq.~(\ref{fitSHVF}) with $A_V = 0.57$, $\beta = 2.6$, and $B=7.0$,
  which roughly describes the average of all SHVFs, and is shown to
  facilitate a comparison.}
\label{fig:Vmax}
\end{figure*}

Comparing panels in different rows, there is no indication for a
significant dependence on host halo mass. However, comparing results
in different columns, different simulations and/or subhalo finders do
seem to result in $V_{\rm max}/V_{\rm acc} - m/m_{\rm acc}$ relations
that are slightly different. Overall, in the case of \Rockstar, the
best-fit relations are in good agreement with the results of Paper~I
and with the simple power-law relation of Hayashi \etal (2003).  This
is not entirely unexpected, given that the Rhapsody project also used
\Rockstar to identify haloes and subhaloes. There is some indication,
though, that the Bolshoi simulation reveals slightly more scatter in
the $V_{\rm max}/V_{\rm acc} - m/m_{\rm acc}$ relation than the
MultiDark simulation. More dramatic are the differences between the
\BDM and \Rockstar analyses of the Bolshoi simulation. The \BDM
results are offset to larger $V_{\rm max}/V_{\rm acc}$ at given
$m/m_{\rm acc}$, and display a larger scatter especially at low
$m/m_{\rm acc}$.

These differences emphasize that not only the abundances, but also the
structural properties of dark matter subhaloes, are sensitive to the
subhalo finder used. Overall, though, the best-fit relations,
indicated by the dashed, red lines, are nicely bracketed by the
results based on idealized simulations by Hayashi \etal (2003) and
Pe\~narrubia \etal (2010), and in general the solid, red line, which
is used in our model, is a reasonable description of the average
trend.  More detailed investigations are needed to investigate the
origin of the differences between the various simulation results,
although we suspect that they are largely due to subtleties related to
how subhaloes masses are determined.
\begin{figure*}
\centerline{\psfig{figure=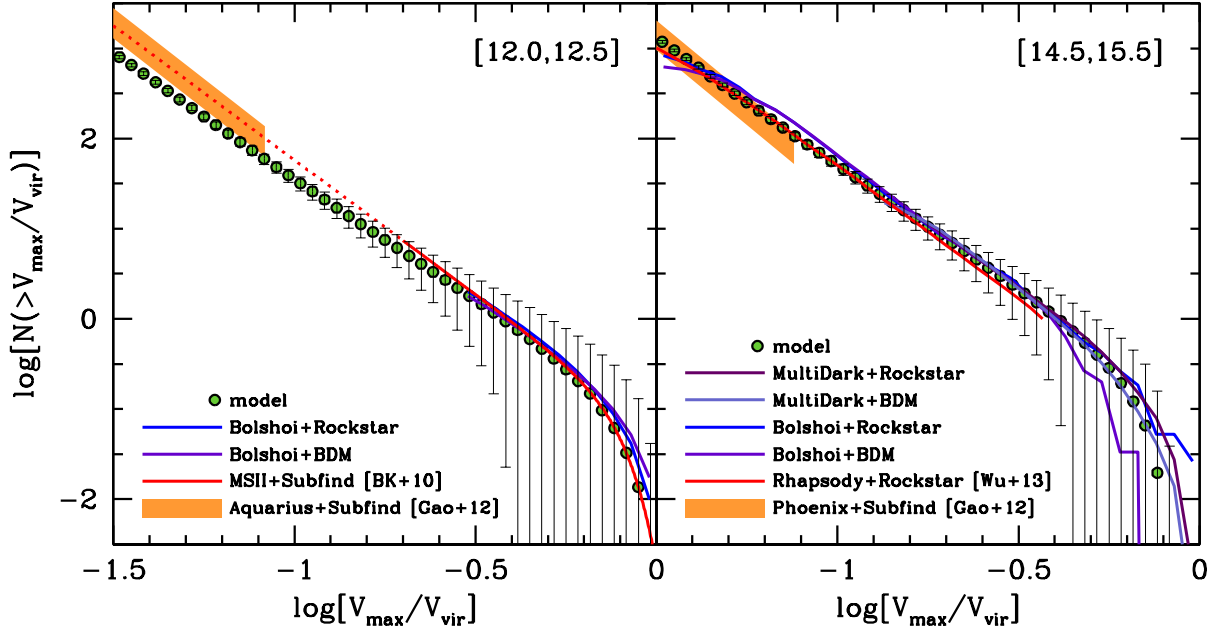,width=0.9\hdsize}}
\caption{Cumulative subhalo velocity functions for host haloes with
  $\log[M/(h^{-1}\Msun)]$ in the range $[12.0,12.5]$ (left-hand panel)
  and $[14.5,15.5]$ (right-hand panel).  The green dots are the
  results obtained using our semi-analytical model, averaging over 5,000
  host haloes, while the errorbars indicate the standard deviation due
  to halo-to-halo variance. In addition to the results from Bolshoi
  and MultiDark (using only subhaloes resolved with at least 250
  particles), we also overplot results from \Rockstar for the Rhapsody
  simulations, and from \Subfind for the Millennium Simulations II, the
  Aquarius simulations and the Phoenix simulations.}
\label{fig:Vmaxcum}
\end{figure*}

\subsection{Subhalo Velocity Functions}
\label{sec:SHVF}

The right-hand panels of Fig.~\ref{fig:Vmax} show the subhalo velocity
functions (SHVFs), $\rmd N/\rmd\log(V_{\rm max}/V_{\rm vir})$,
obtained using our semi-analytical model. These correspond to the SHMFs
shown in the corresponding panels in Fig.~\ref{fig:SHMF}, and reveal a
similar dependence of normalization on host halo mass, at least at the
low velocity end. In particular, the model predicts that more massive
host haloes have a larger abundance of subhaloes at fixed $V_{\rm
  max}/V_{\rm vir} \lta 0.6$.  At the high velocity end, though, where
the velocity function cuts off exponentially, this trend reverses
sign. This indicates that, unlike the SHMFs, the SHVFs are not
self-similar! As shown in Paper~I, this is already imprinted in the
unevolved SHVF (i.e., using the values of $V_{\rm max}$ at accretion),
and is a consequence of the concentration-mass-redshift relation of
dark matter haloes, which induces a mass dependence in the ratio
$V_{\rm max}/V_{\rm vir}$ of host haloes.  The model SHVFs are well
fit by
\begin{eqnarray}\label{fitSHVF}
\lefteqn{{\rmd N \over \rmd\log(V_{\rm max}/V_{\rm vir})} = } \nonumber \\
& & A_V \, \left({V_{\rm max} \over V_{\rm vir}}\right)^{-\beta} \,
\exp\left[-B\,(V_{\rm max}/V_{\rm vir})^{15}\right]\,.
\end{eqnarray}
with slope $\beta = 2.6$ and a normalization $A_V$ that depends
(weakly) on host halo mass. Contrary to the SHMFs, the scale parameter
$B$ now also depends on halo mass (and, as discussed in detail in
Paper~I, also on redshift and cosmology). The green, dashed line in
each panel corresponds to Eq.~(\ref{fitSHVF}) with $A_V = 0.57$,
$\beta = 2.6$, and $B=7.0$, which roughly describes the average of all
SHVFs shown.

The left and middle panels of Fig.~\ref{fig:Vmax} show the SHVFs
obtained from the MultiDark (upper panels) and Bolshoi (lower panels)
simulations using both \BDM (left-hand panels) and \Rockstar (middle
panels). The solid curves are the SHVFs obtained using only subhaloes
with at least 250 particles. The dotted parts of the curves show the
extensions one obtains when including subhaloes down to a limit of 50
particles per subhalo, which is the same limit as used for the SHMFs
in Fig.~\ref{fig:SHMF}. As is evident, including haloes with fewer
than $\sim 250$ particles results in SHVFs that are clearly affected
by resolution effects. Boylan-Kolchin \etal (2010), in their analysis
of the Millennium II simulation, argued that the minimum number of
particles required to resolve haloes well enough for a reliable
estimate of $V_{\rm max}$ is 150.  Our results, though, indicate that
this limit has to be increased to $\sim 250$, at least for the
MultiDark and Bolshoi simulations used here.

Comparing the SHVFs obtained from the simulations with those predicted
by our semi-analytical model, similar trends are apparent as for the SHMFs.
Simulation results and model predictions are in good agreement when it
comes to the normalization and slope at the low-velocity end of the
SHVFs (i.e., where $V_{\rm max}/V_{\rm vir} \lta 0.5$). However, at
the massive end, where the SHVFs reveal an exponential cut-off,
significant differences are apparent, albeit less pronounced than in
the case of the SHMFs. As with the subhalo mass functions, the \BDM
results reveal a much more pronounced dependence on host halo mass
compared to either the \Rockstar results or the predictions from our
semi-analytical model. This most likely is yet another manifestation of the
problem that \BDM has with the treatment of massive subhaloes (see
discussion in \S\ref{sec:BDM}).

Fig.~\ref{fig:Vmaxcum} compares the {\it cumulative} subhalo velocity
functions, $N(>V_{\rm max}/V_{\rm vir})$ for two bins in host halo
mass: $\log[M/(h^{-1}\Msun)] \in [12.0,12.5]$ (left-hand panel) and
$[14.5,15.5]$ (right-hand panel). The green dots are the results
obtained using our semi-analytical model, averaging over 5,000 host haloes,
while the errorbars indicate the standard deviation due to
halo-to-halo variance. In addition to the results from Bolshoi and
MultiDark (using only subhaloes resolved with at least 250 particles),
we also overplot results from \Rockstar for the Rhapsody simulations,
and from \Subfind for the Millennium Simulations II, the Aquarius
simulations and the Phoenix simulations. The results from the Aquarius
and Phoenix simulations are taken from Gao \etal (2012), who defined
host haloes as spheres with a radius of $r_{200c}$. We have converted
their results to our definition of halo mass and radius. The lower
boundary of the orange band shown corresponds to a conversion in which
we assume that subhaloes follow the density distribution of the dark
matter, while the upper boundary marks the results obtained assuming
that subhaloes are homogeneously distributed within their host. These
two extremes bound the true distribution of subhaloes.  Overall the
agreement between the various simulation results and the semi-analytical
model is extremely good.  In particular, the MSII + \Subfind results
from Boylan-Kolchin \etal (2010) accurately match the model
predictions as well as both the \Rockstar and \BDM results obtained
from the Bolshoi simulation. This demonstrates that the dramatic
discrepancies between these results in the SHMFs
(cf. Fig.~\ref{fig:comparison}) arise from issues related to assigning
masses to the subhaloes, rather than from issues related to detecting
them. The maximum circular velocity probes the inner regions of dark
matter haloes, and is therefore a much more robust quantity to measure
in simulations than subhalo mass.
\begin{figure*}
\centerline{\psfig{figure=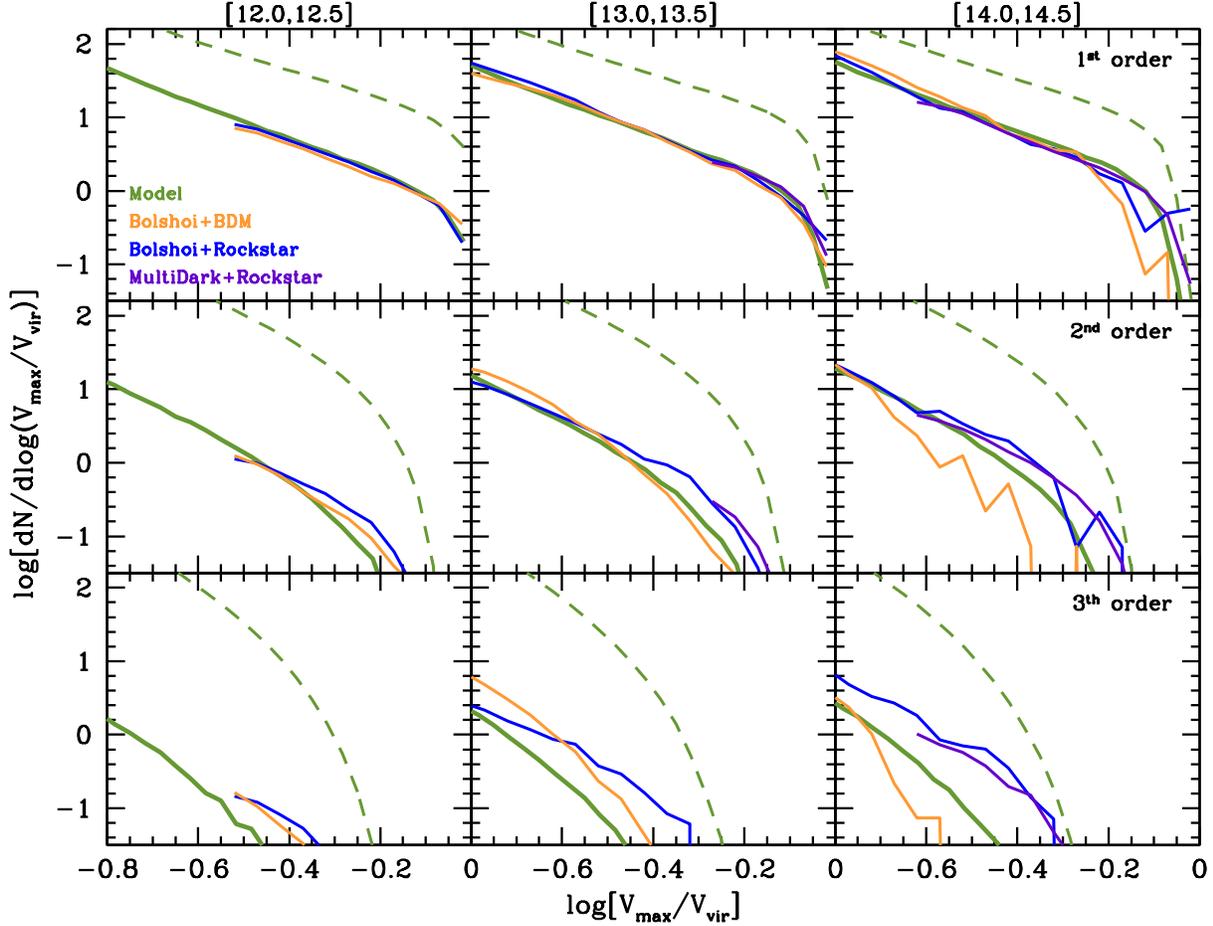,width=0.9\hdsize}}
\caption{Subhalo velocity functions of first order (upper row), second
  order (sub-subhaloes, middle row) and third order
  (sub-sub-subhaloes; lower row). Results are shown for three
  different bins in host halo mass, as indicated by the values in
  square brackets at the top of each column. Solid, green lines are
  the model results, obtained by averaging over 5,000 host haloes. The
  other curves are the results obtained from the Bolshoi and MultiDark
  simulations, using only subhaloes with at least 250 particles, as
  indicated in the top-left panel. The dashed, green curves are the
  unevolved SHVFs, as obtained from our semi-analytical model, and are
  shown for comparison.}
\label{fig:orders}
\end{figure*}

The main deviant with respect to our model predictions are the
Aquarius results of Gao \etal (2012), which have a larger
normalization and a significantly steeper slope. Note that the
extrapolation of the MSII + \Subfind results by Boylan-Kolchin \etal
(2010), indicated by the red, dotted curve in the left-hand panel, is
in good agreement with the Aquarius results, suggesting that this is
largely a \Subfind issue. Gao \etal (2012) fitted their cumulative
SHVFs with a power-law, $N(>V_{\rm max}) \propto V_{\rm max}^{-\zeta}$,
over the range $0.025 \leq V_{\rm max}/V_{200c} \leq 0.1$, which
roughly corresponds to $0.02 \leq V_{\rm max}/V_{\rm vir} \leq 0.08$,
and measured slopes of $\zeta = 3.13$ and $3.32$ for the Aquarius and
Phoenix simulations, respectively. This is significantly steeper than
our model predictions, which have $\zeta \simeq 2.75\pm
0.05$\footnote{We caution that, depending on the range over which it
  is measured, the slope of the cumulative velocity function, $\zeta$,
  can differ significantly from that of the differential SHVF,
  $\beta$}.  Unfortunately, we have not been able to find any other
simulation results that can be used to investigate these discrepancies
for the $[12.0,12.5]$ mass bin. However, in the case of the
cluster-sized haloes, we can include the Rhapsody and Bolshoi
simulations in the comparison. Whereas our model predictions are in
excellent agreement with the Rhapsody results, they are somewhat too
low around $V_{\rm max}/V_{\rm vir} \sim 0.1$ compared to the Bolshoi
results. Given the level of disagreement between the various
simulation results, we conclude that, at this stage, there is no
indication that the model predictions are incorrect. More detailed
simulation results are required to further test our model prediction
that the slope of the (cumulative) SHVF is significantly shallower
than 3.0. We emphasize that accurate knowledge of this slope is
important for various areas of astrophysics, in particular for making
accurate predictions for the expected dark matter annihilation signal
(e.g., Bergstrom \etal 1999; Colafrancesco, Profumo \& Ullio 2006;
Giocoli \etal 2008b).

\subsection{Higher-order substructure}
\label{sec:highorder}

Thus far, all subhalo mass and velocity functions shown are for {\it
  all} subhaloes, irrespective of their order. We now shift our focus
to subhaloes of different orders. In what follows we only present
results for subhalo velocity functions, though results for the subhalo
mass functions are qualitatively similar. Fig.~\ref{fig:orders} plots
the SHVFs for subhaloes of first, second and third order (different
rows) for host haloes in three different bins of host halo mass, as
indicated at the top of each column. Solid green lines are the model
predictions, obtained averaging over 5,000 host haloes.  The dashed,
green lines are the corresponding {\it unevolved} subhalo velocity
functions (i.e., using $V_{\rm acc}$ rather than $V_{\rm max}$) of the
same order. The evolved SHVFs are reduced with respect to the
unevolved ones due to mass stripping, which reduces the subhalo's
$V_{\rm max}$ according to Eq.~(\ref{Vmaxfit}).

Overplotted are the results obtained from the Bolshoi and MultiDark
simulations. They are in superb agreement with our model predictions
for the first-order subhaloes. However, very significant discrepancies
are apparent for the second- and third-order subhaloes, not only
between the simulation results and our model predictions, but also
between the different simulation results. Whereas the \Rockstar
results obtained from Bolshoi are in good agreement with those from
MultiDark, they are very different from those obtained using \BDM,
especially in the more massive host haloes. These discrepancies among
the simulation results make it difficult to judge the reliability of
the model. In principle, there are a few oversimplifications in the
model that could result in inaccuracies. First of all, in our model
subhaloes can only increase their order with time. This ignores the
possibility that higher-order subhaloes are stripped from their direct
parent, thereby reducing their order by one.  If this occurs frequently,
our model will underestimate the abundance of subhaloes of a given
order. Our model also ignores potential mergers among subhaloes.
However, since there is no obvious reason for stripping or merging to
be more frequent or relevant for higher-order subhaloes than for
first-order subhaloes, and given the fact that our model is in
excellent agreement with the simulation results for first-order
subhaloes, we consider it unlikely that these shortcomings would have
a significant impact. Rather, we believe that these discrepancies are
more likely a manifestation of the subtle differences between how
model and simulations treat `overlapping' haloes (see \S\ref{sec:massdef}
and Fig.~\ref{fig:illustration}).

\section{Summary}
\label{sec:conc}

We have compared subhalo mass and velocity functions obtained from
different simulations, with different subhalo finders, among each
other and with predictions from our new semi-analytical model presented in
Paper~I. Our findings can be summarized as follows:
\begin{itemize}

\item We confirm the findings of Onions \etal (2012) and Knebe \etal
  (2013) that the subhalo mass functions obtained using different
  subhalo finders agree with each other at the level of $\sim 20$
  percent. However, this is only true for low-mass subhaloes with $m/M
  \lta 0.1$; at the more massive end, different subhalo finders yield
  SHMFs that differ by more than an order of magnitude!

\item Subhalo finders that identify subhaloes based purely on density
  in configuration space, such as \Subfind and \BDM, dramatically
  underpredict, by more than an order of magnitude, the abundances of
  massive subhaloes (with masses $m \gta M/10$), especially in more
  massive host haloes. These problems are much less severe for the
  subhalo velocity functions, indicating that they arise from issues
  related to assigning masses to the subhaloes, rather than from
  issues related to detecting them. The maximum circular velocity
  probes the inner regions of dark matter haloes, and can therefore be
  measured much more reliably than subhalo mass.

\item Overall our model predictions are in excellent agreement with
  simulation results obtained using the more advanced subhalo finders
  \Rockstar and \SURV. In particular, the model accurately reproduces
  the slope and host-mass-dependent normalization of both the subhalo
  mass and velocity functions.  There are small discrepancies at the
  very massive end, but rather than reflecting an inaccuracy of the
  model, these arise from subtle issues having to do with the exact
  halo mass definitions of overlapping haloes.

\item Since tidal stripping and heating impact the outskirts of
  subhaloes much more than their inner regions, a large reduction in
  mass only has a relatively mild impact on the maximum circular
  velocity (Hayashi \etal 2003; Pe\~narrubia \etal 2008, 2010). We
  confirm our findings from Paper~I that, on average, the relation
  between $V_{\rm max}/V_{\rm acc}$ and $m/m_{\rm acc}$ is well
  described by Eq.~(\ref{Vmaxfit}) with $(\eta,\mu) = (0.44,0.60)$,
  which is roughly bracketed by the relations obtained by Hayashi
  \etal (2003) and Pe\~narrubia \etal (2010) using idealized $N$-body
  simulations.  However, there are small but noticeable differences in
  the best-fit values of $\eta$ and $\mu$ for different subhalo
  finders, indicating that not only the abundances, but also the
  structural properties of dark matter subhaloes, are sensitive to the
  subhalo finder used.

\item The mass and velocity functions obtained from the Bolshoi and
  MultiDark simulations confirm our finding from Paper~I that the
  power-law slopes of $\rmd N/\rmd\log(m/M)$ and $\rmd
  N/\rmd\log(V_{\rm max}/V_{\rm vir})$ are with $0.82$ and $2.6$,
  respectively, significantly shallower than what has been claimed in
  several studies in the literature. In particular, studies based on
  \Subfind by Boylan-Kolchin \etal (2010) and Gao \etal (2012) have
  yielded slopes that are significantly steeper. Given the excellent
  agreement between our model predictions and the \Rockstar, \BDM and
  \SURV results, we believe that this discrepancy reflects a problem
  with \Subfind.  We emphasize that accurate knowledge of the
  power-law slope of the subhalo mass and velocity functions is
  important for calculating the `boost' factor for dark matter
  annihilation due to substructure, as this requires extrapolation of
  $\rmd N/\rmd\log(m/M)$ down to the dark matter Jeans mass.

\item Comparing the velocity functions for subhaloes of different
  order, we find that our model is in excellent agreement with the
  Bolshoi and MultiDark results for first-order subhaloes. This
  agreement, however, rapidly deteriorates with increasing order; not
  only between model and simulations, but also among the simulation
  results themselves. We speculate that these discrepancies are mainly
  a manifestation of subtle issues having to do with how different
  subhalo finders treat overlap among haloes and subhaloes. More
  detailed studies are required to investigate these issues further,
  and to provide a more reliable testbed for our model predictions.

\end{itemize}

\section*{Acknowledgments}

We are grateful to the people responsible for the Bolshoi and
MultiDark simulations for making their halo catalogs publicly
available, and to the following individuals for their advice and
assistance: Peter Behroozi, Mike Boylan-Kolchin, Carlo Giocoli, Andrew
Hearin, Anatoly Klypin, Alexander Knebe, Surhud More, Nikhil
Padmanabhan, Tomomi Sunayama, and Bepi Tormen.


\label{lastpage}

\end{document}